# Solution to the Twin Image Problem in Holography

Tatiana Latychevskaia & Hans-Werner Fink

*Institute of Physics, University of Zurich, Winterthurerstrasse 190, CH-8057, Switzerland*

**While the invention of holography by Dennis Gabor[1,2] truly constitutes an ingenious concept, it has ever since been troubled by the so called twin image problem limiting the information that can be obtained from a holographic record. Due to symmetry reasons there are always two images appearing in the reconstruction process. Thus, the reconstructed object is obscured by its unwanted out of focus twin image. Especially for emission electron as well as for x- and gamma-ray[3,4] holography, where the source-object distances are small, the reconstructed images of atoms are very close to their twin images from which they can hardly be distinguished. In some particular instances only, experimental efforts could remove the twin images[5,6,7]. More recently, numerical methods to diminish the effect of the twin image have been proposed but are limited to purely absorbing objects[8,9,10,11,12] failing to account for phase shifts caused by the object. Here we show a universal method to reconstruct a hologram completely free of twin images disturbance while no assumptions about the object need to be imposed. Both, amplitude and true phase distributions are retrieved without distortion.**

The basic setup for holography is depicted in Figure 1 together with the positions of the object and its twin image when an in-line hologram is reconstructed. In the object plane, the twin appears as an out of focus image while in the twin image plane the object appears out of focus. The two images are mirror-symmetric with respect to the point source. In holography with visible light, the object and its twin image can be separated by using parallel beams and subtracting a second hologram from the reconstructed image[1,5], by employing a beam splitter[6] or introducing additional lenses into the recording and reconstructing scheme[7]. However, lenses are not available for x-ray or gamma-ray holography. In electron emission holography, the close proximity of source and sample also makes it impossible to employ lenses or a beam-splitter between them. In other schemes, like holography with low energy electrons lenses are to be avoided due to their inherent aberrations. Moreover, in-line holography exhibits high phase sensitivity and is therefore, for coherent low energy electrons[13] and even for high energy electrons[14], for which DNA molecules represent extremely weak phase objects, the method of choice.

The most widely employed approach to address the twin image problem is to record a set of holograms at different wavelengths[15,16]. However, this method only suppresses but not eliminates the twin image and is experimentally difficult to implement in particular when it comes to record fragile biological molecules subject to radiation damage. So far numerical methods to diminish the effect of the twin image have been restricted to holograms of purely absorbing objects[8,9,10,11,12], a coarse approximation of physical reality. In this letter we show how the twin image can be eliminated by numerical reconstruction of a hologram without imposing any restrictions on or assumptions about the object to be imaged.

A reference wave $A\exp(ikr)$ where $A$ is a complex constant and $\boldsymbol{r}$ is the radius vector to some point in space, propagates from a point source towards a distant screen illuminating it with the intensity $|A\exp(ikr_s)|^2 = |A|^2 = B$, where $\boldsymbol{r}_s$ describes a point on the screen, thus providing a coherent background $B$. If an object is placed into the beam, the hologram is formed at the screen. It is brought about by interference between the wave scattered by the object and the un-scattered wave. The transmission function in the object plane can be written as $(1+t(\boldsymbol{r_0}))$, where 1 corresponds to the transmittance in the absence of the object, $t(\boldsymbol{r_0})$ is a complex function describing the presence of the object and $\boldsymbol{r_0}$ describes a point of the object. Part of the beam passes the object un-scattered forming the reference wave, $A\exp(ikr_0)$. The part of the beam scattered by the object gives rise to the object wave, $A\exp(ikr_0)t(\boldsymbol{r_0})$. The total field at the screen is the sum of the reference and object wave $A(R_0(\boldsymbol{r}_s)+O_0(\boldsymbol{r}_s))$, where $R_0=\exp(ikr_s)$, and $AO_0(\boldsymbol{r}_s)$ is the object wave distribution on the screen, which is calculated by solving the Kirchhoff-Helmholtz integral[17] $AO_0(\boldsymbol{r}_s) = \iint A\exp(ikr_0)t(\boldsymbol{r}_0)\exp(-ikr_0\boldsymbol{r}_s/r_s)d\sigma_0$, where $\sigma_0$ denotes the object plane. The interference pattern on the screen can be recorded by a sensitive medium, yielding a hologram with the transmission function $H(\boldsymbol{r}_s) = |A|^2|R_0(\boldsymbol{r}_s)+O_0(\boldsymbol{r}_s)|^2$. Dividing the hologram image by the background image results in $H(\boldsymbol{r}_s)/B(\boldsymbol{r}_s) = |R_0(\boldsymbol{r}_s)+O_0(\boldsymbol{r}_s)|^2$ which we call the normalized hologram. It is worth noting that this normalized hologram is independent of $|A|^2$. This is an important experimental aspect since $|A|^2$ would vary with changes in the point source intensity, camera sensitivity, image intensity-scale defined by the image format, etc. The following routine is applied to such normalized holograms making it independent of details of the data acquisition.

The final goal of our method is to reconstruct the distribution of the complex sum $(R_0(r_s)+O_0(r_s))$. This is achieved by an iterative procedure[18,19,20] which basically boils down to the field propagation back and forth between the screen- and the object-plane, until all artefacts due to the twin image are gone. It includes the following steps:

(i) Formation of the input complex field as $(R_0(r_s)+O_0(r_s)) = |R_0(r_s)+O_0(r_s)|\exp(i\Omega(r_s))$ where the amplitude is always given by the square root of the normalized hologram $|R_0(r_s)+O_0(r_s)| = \sqrt{H(r_s)/B(r_s)}$, and the phase $\Omega(r_s)$ is initially set to $kr_s$ - the phase of the known reference wave $R_0=\exp(ikr_s)$ – and it evolves towards its true value during iteration.

(ii) Back propagation to the object plane is simulated using the Helmholtz-Kirchhoff formula[17].

(iii) The reconstructed complex field distribution multiplied with the conjugated incident wave $\exp(-ikr_0)$ gives the complex transmission function in the object plane $(1+t(r_0))$. The extracted complex transmission function describes the object's absorption and phase shift by the relation: $1+t(r_0) = (1-a(r_0))\exp(-i\varphi(r_0))$, where $a(r_0)$ defines the absorbing properties of the object and $\varphi(r_0)$ the phase shift introduced by the object with respect to the incident wave. Thus, due to the presence of the reference wave, the correct absorption $a(r_0)$ and phase properties $\varphi(r_0)$ of the object can be extracted. The following constraint can be applied: $a(r_0) \geq 0$, which refers to nothing else but the basic physical notion of energy conservation requiring that absorption may not lead to an increased amplitude following a scattering process. This in turn implies that whenever negative values of $a(r_0)$ emerge, they are the result of the interference between the twin image- and the reference-wave, and are subsequently replaced by zeros while the phase values remain unchanged. Purely based on this basic physical notion of positive absorption values, we obtain a constraint to derive a re-combined absorption and phase distribution for the object leading to the new transmission function:

$$1 + t'(r_0) = (1 - a'(r_0))\exp(-i\varphi(r_0)).$$

(iv) Next, we let the reference wave $A\exp(ikr)$, originating from the point source, propagate[17] forward. Once it has passed the object with the new transmission function $(1 + t'(r_0))$ we arrive at the new complex sum $(R'_0(r_s) + O'_0(r_s))$ in the screen plane. We then acquire the altered phase value from this new sum and use it as the input phase value for the next iteration starting at step (i). We would like to point out that the first iteration already reconstructs the complex object with the same quality as the conventional hologram reconstruction routines. Further iterations eventually lead to the elimination of the twin term. There are **no** limitation on the object surrounding, object's size or on the object properties being weak absorbing or weak/strong phase shifting. The object and its surrounding can be anything: for instance, an extended biological molecule which is identified only by a weak phase shift of the incident beam, or an arrangement of heavy atoms acting as point-like absorbing centres.

We now test the routine, first with a simulated hologram, then by using experimental holograms. A hologram of an extended object with a maximum absorption $a(r_0)$ of 80% of an incident beam and a maximum phase shift $\varphi(r_0)$ of 3 radians was simulated by using the Helmholtz-Kirchhoff formula[17] and reconstructed with our method described above. The results are shown in Figure 2. The object's absorption and phase distributions obtained by conventional hologram reconstruction routine are shown in Figure 2b, and those obtained after the first iteration in Figure 2c. Both reconstructions are similar and suffer from the superimposed out of focus twin image. While the reconstructed absorption distributions are almost identical, the phase distribution obtained by the conventional reconstruction is initially better since it shows a maximum phase shift of 2.5 radian (see Figure 2b). Nevertheless, it does not reach the correct pre-defined maximum phase shift of 3 radian. Each further iteration step causes the twin image to fade away in the reconstructed absorption distribution while the phase distribution approaches its true values. After the 1-st iteration step the maximal phase shift amounts only 0.2 radian (see Figure 2c). However, already after the 10-th iteration (see Figure 2d) the phase distribution appears almost free from the disturbing twin image and the phase values are recovering. The absorption and phase distributions, retrieved after the 50-th iteration, shown in Figure 2e, demonstrate that the effect of the twin image has been completely removed and the phase has finally reached its pre-defined value of 3 radians. This demonstrates that the twin image removal is more than just getting rid of an artefact of holography; it is also a way to arrive at the true phase values.

To apply our method to experimental holograms, green laser light of 532 nm wavelength (Changchun New Industries Optoelectronics diode laser) is focused by means of a microscope objective (Newport M60, NA=0.85). At the focal point a pinhole of 20 μm in diameter optimises the beam to form a coherent point source. As an object a tungsten tip is then placed into the divergent beam and controlled by a xyz-movable stage. The distance between the point source and the tip is adjusted to approximately 0.81 mm. A screen is placed at about 1 m from the point source and the hologram captured by a CCD camera (Hamamatsu C4742-95). The recorded and normalized hologram and its reconstructions are shown in Figure 3. The result after 500 iterations shows that residues due to the twin image in the reconstructed absorption and phase distributions are gone.

With this, a novel method to finally solve the twin image problem is established and can now be applied without limitations to wavelength or wave front shapes (planar, or spherical), for imaging objects of arbitrary size, exhibiting absorbing and/or phase shifting properties. From a single holographic record, twin-image free true absorption and phase distributions are iteratively retrieved.

**Acknowledgements:** We would like to thank Conrad Escher for his help in preparing the experimental hologram samples and for carefully reading the manuscript. The work presented here is supported by the Project SIBMAR, part of the "New and Emerging Science and Technology" European Program.

**Competing interest statement** The authors declare that they have no competing financial interests.

**Correspondence** and requests for materials should be addressed to Tatiana Latychevskaia (tatiana@physik.unizh.ch)

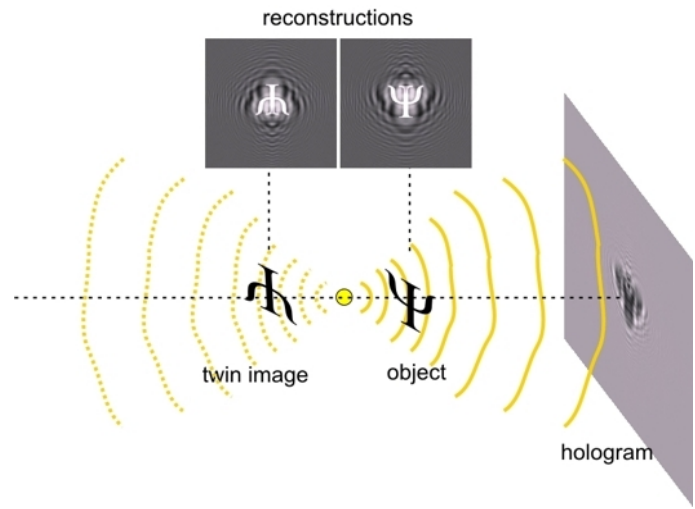

**Figure 1: Position of the object and its twin image during hologram reconstruction.**

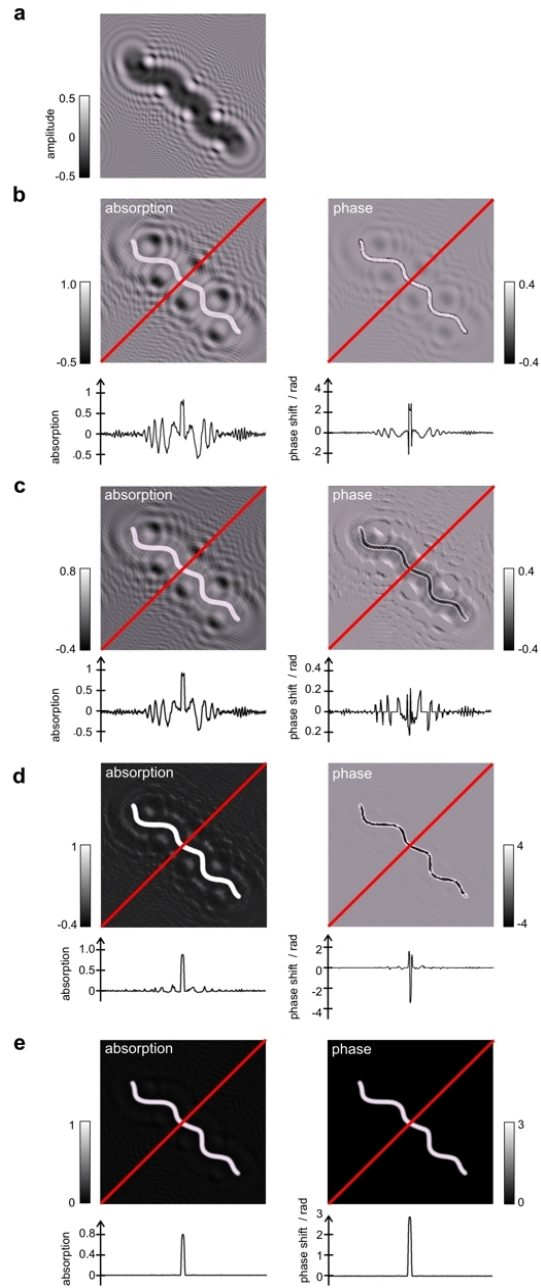

**Figure 2: Simulated and reconstructed hologram of an extended object. a**, Normalized hologram. **b**, Reconstructed amplitude and phase by conventional reconstruction. **c**, Reconstructed amplitude and phase distributions after the first iteration. **d**, Reconstructed amplitude and phase distributions after the 10-th iteration. **e**, Reconstructed amplitude and phase distributions after the 50-th iteration. Below each reconstruction the intensity distributions along the cuts shown in red are displayed.

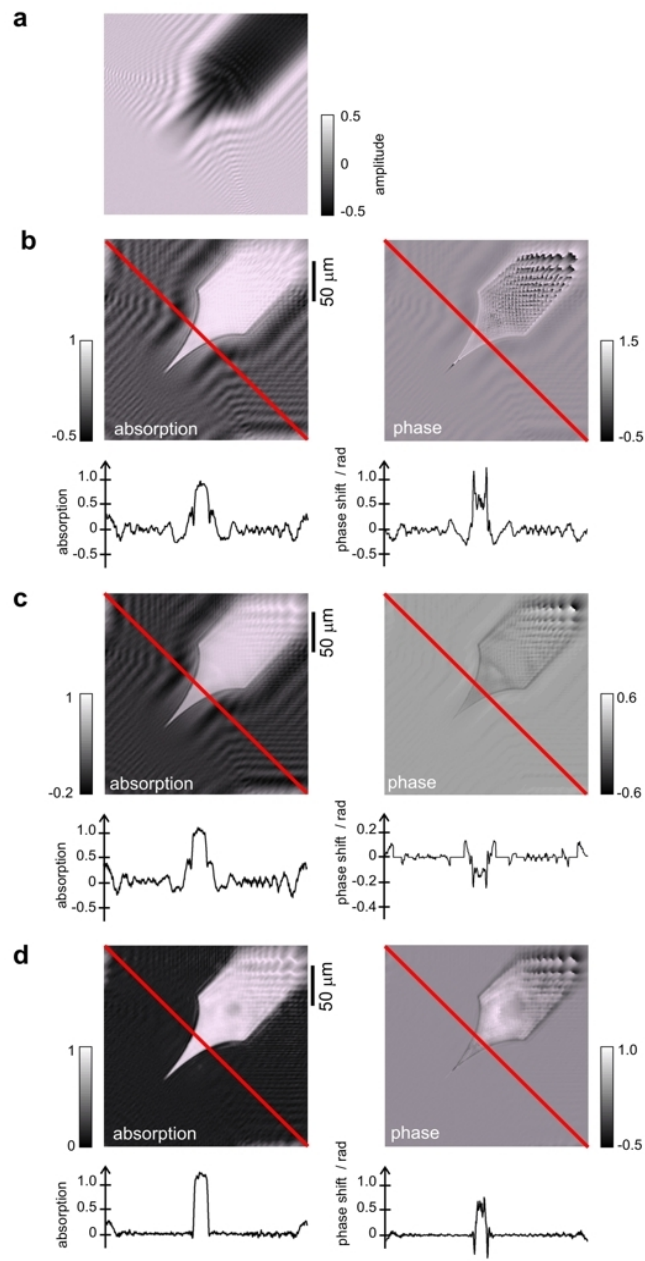

**Figure 3: Iteratively reconstructed experimental optical hologram of a tungsten tip. a**, Normalized hologram. **b,** Reconstructed absorption and phase distributions by conventional reconstruction. **c,** Reconstructed absorption and phase distributions after the first iteration. The oscillations due to the twin image are apparent. **d,** Reconstructed amplitude and phase distributions after the 500-th iteration. Below each reconstruction the intensity distributions along the red lines are displayed.